\documentstyle[preprint,aps,prb]{revtex}
\begin{document}
\draft
%\twocolumn[\hsize\textwidth\columnwidth\hsize\csname %
%@twocolumnfalse\endcsname

\title{Normal Fermi Liquid Behavior of Quasiholes in the Spin--Polaron
Model
for Copper Oxides }
\author{ G. Jackeli and V. Yu. Yushankhai }
\address{Joint Institute for Nuclear Research, Dubna 141980, Russia }
\date{\today}
\maketitle
\begin{abstract}\widetext
Based on the $t-J$ model and the self--consistent Born approximation,
the damping of quasiparticle hole states near the Fermi surface is
calculated in a low doping regime. Renormalization of spin--wave
excitations due to hole doping is taken into account. The damping
is shown to be described by a familiar form
$\text{Im}\Sigma({\bf k}^{\prime},\varepsilon)\propto
(\varepsilon^{2}/ \varepsilon_{F})\ln(\varepsilon/ \varepsilon_{F})$
characteristic of the 2--dimensional Fermi liquid, in contrast with the
earlier statement reported by Li and Gong
[Phys. Rev. B {\bf 51}, 6343 (1995)]
on the marginal Fermi liquid behavior of quasiholes.

\end{abstract}

\pacs{PACS numbers: 71.27+a, 72.10.Di, 72.20.Jv, 75.10Jm}
%]
\narrowtext
%\newpage

Understanding of the quasiparticle (QP) characteristics of
charge carriers forming the normal--state electronic properties
of high--$T_{c}$ superconductors (HTSC's)
is an issue of current interest.\cite{BR}
In particular, it is now under debate whether
these compounds can be described within
the normal Fermi liquid (FL) approach, or a more exotic scenario,
for instance, the marginal FL (MFL) concept,\cite{MFL}
should be involved.

In an attempt to understand the QP properties of HTSC's
one has to take into account a strong difference between
an intermediate and a low doping regime as it follows from
the angle resolved photoemission (ARPES) experiments.\cite{AR}
Actually, at the intermediate level of doping ARPES indicates a
large Fermi surface (FS), while the reference
insulator compounds \cite{W} show a hole dispersion that is
compatible with a small four--pocket shape of FS at low doping.
In the latter regime, a hole propagation is strongly affected by the
presence of antiferromagnetic (AFM) correlations in the spin subsystem.
The essential features of this problem are described by the $t-J$ model.

In the present paper, we investigate the regime
of low doping based on the $t-J$ model in the slave--fermion
Schwinger boson representation. We are mainly interested in a
QP hole behavior near FS.
We show that at zero temperature the imaginary part of the hole
self--energy $\text{Im}\Sigma ({\bf k}, \omega)\propto
\omega^{2} \ln \omega$, which indicates a conventional FL
behavior of quasiholes. Our result is at variance with
 one reported in Ref.\ \onlinecite{Kit}, where the MFL--behavior
of quasiholes is obtained even at $T=0$.
The reason of this contradiction is discussed below.

 By using the slave--fermion Schwinger boson factorization
 for electron operators, the $t-J$ model with the Neel ground state
can be mapped onto the so--called spin--polaron \cite{V,KLR,MH,IF}
model with the Hamiltonian given by
\begin{eqnarray}
H&=&\sum_{{\bf q}}^{}\omega_{{\bf q}}\alpha_{{\bf q}}^{\dagger}
\alpha_{{\bf q}}
- \mu \sum_{{\bf k}} h_{{\bf k}}^{\dagger}h_{{\bf k}} \nonumber \\
 &&+\frac{zt}{\sqrt{N}} \sum_{{\bf k},{\bf q}}^{}
h_{{\bf k}}^{\dagger}h_{{\bf k}-{\bf q}}
[M_{1}({\bf k},{\bf q})\alpha_{{\bf q}}+
M_{2}({\bf k},{\bf q}) \alpha^{\dagger}_{-{\bf q}}],
\label{int}
\end{eqnarray}
where $\omega_{{\bf q}}=zJ/2 \sqrt{1-\gamma_{{\bf q}}^{2}}$
with $\gamma_{{\bf k}}=\frac{1}{2}(\cos k_{x}+\cos k_{y})$,
$\mu$ is the chemical potential of holes,
$M_{1}({\bf k},{\bf q})= M_{2}({\bf k}-{\bf q},-{\bf q})
=(u_{{\bf q}}\gamma_{{\bf k}-{\bf q}}+v_{{\bf q}}\gamma_{{\bf k}})$,
and $z=4$ for the square lattice;
$N$ is the total number of sites, the
lattice spacing is taken to be unity, and $u_{{\bf q}}$,
$v_{{\bf q}}$ are the usual parameters
of the Bogoliubov $u-v$ transformation.
In the Hamiltonian (\ref{int}) $h_{{\bf k}}$
($h_{{\bf k}}^{\dagger}$) are canonical spinless
fermion operators and $\alpha_{{\bf q}}$
($\alpha_{{\bf q}}^{\dagger}$) are canonical boson
operators.

We introduce the Fourier transformed two--time retarded Green
function (GF) $G({\bf k},\omega)=\ll h_{{\bf k}}|
h_{{\bf k}}^{\dagger}\gg_{\omega}$
for fermions and the matrix GF
 $D({\bf q},\omega)=
 \ll A_{{\bf q}}|A_{{\bf q}}^{\dagger}\gg_{\omega}$
for magnons, $ A_{{\bf q}}$
is the two--component operator and
$A_{{\bf q}}^{\dagger} = ( \alpha_{{\bf q}}^{\dagger}, \;
\alpha_{- {\bf q}} )$.
By applying the irreducible GF method \cite{TS} and
using a decoupling procedure, which is equivalent to
the self--consistent Born approximation (SCBA), both for
$G({\bf k},\omega)$ and $D({\bf q},\omega)$,
 we obtain
\begin{equation}
G^{-1}({\bf k},\omega)=
\omega+\mu-\Sigma({\bf k},\omega),
\label{G}
\end{equation}
%\widetext
\begin{eqnarray*}
 D({\bf q},\omega)&=&
\frac{1}{{\cal D}_{{\bf q}}(\omega)}\nonumber \\
&&\times\!\!\left(\!\begin{array}{cc}
\omega_{{\bf q}}+\omega+\Pi_{22}({\bf q},\omega) &\!
-\Pi_{12}({\bf q},\omega)
\\
-\Pi_{21}({\bf q},\omega)&\!\omega_{{\bf q}}-\omega+
\Pi_{11}({\bf q},\omega)
\end{array}
\!\!\right)
,
\end{eqnarray*}
where
$ {\cal D}_{{\bf q}}(\omega)=[\omega-
\Pi ^{-}({\bf q},\omega)]^{2}
-[\omega_{{\bf q}}+\Pi ^{+}({\bf q},\omega)]^{2} +
\Pi _{12}({\bf q},\omega)\Pi _{21}({\bf q},\omega)$
with
$\Pi^{\pm}({\bf q},\omega)=
1/2[\Pi _{11}({\bf q},\omega)
\pm \Pi_{22}({\bf q},\omega)]$.
 The elements of the polarization operator
$\Pi({\bf q},\omega)$ for the magnon GF has the
form
\narrowtext
\begin{eqnarray}
\Pi_{\alpha \beta}({\bf q},\omega)&=&
\frac{(zt)^{2}}{N}\sum_{{\bf k}}^{}g_{\alpha \beta}({\bf k},{\bf q})
\int\!\!\!\!\int\limits_{-\infty }^{\infty }
d\omega_{1}d\omega_{2}\nonumber \\
&&\times\left[ n(\omega_{2})
-n(\omega_{1})\right]
\frac{\rho_{{\bf k}}(\omega_{1})\rho_{{\bf k}-{\bf q}}(\omega_{2})}
{\omega-\omega_{1} +
\omega_{2}+i\eta}
,
\label{P}
\end{eqnarray}
where $g_{\alpha \beta}({\bf k},{\bf q})=M_{\alpha}({\bf k},{\bf q})
M_{\beta}({\bf k},{\bf q})$ and
$\rho_{{\bf k}}(\omega)=-1/\pi \mbox{Im}G({\bf k},\omega)$ is the
hole spectral function. The
hole self--energy $ \Sigma({\bf k},\omega)$ is given by
\begin{eqnarray}
\Sigma({\bf k},\omega)&=&\frac{(zt)^{2}}{N}
\sum_{{\bf q}}^{}\int\!\!\!\int\limits_{-\infty }^{+\infty }
d\omega_{1}d\omega_{2}
\left[ N(\omega_{2})+1-n(\omega_{1})\right]\nonumber \\
&&\times\frac{\rho_{{\bf k}-{\bf q}}(\omega_{1})
\chi_{{\bf k},{\bf q}}(\omega_{2})}
{\omega-\omega_{1}-
\omega_{2}+i\eta}
,
\label{H}
\end{eqnarray}
where $n(\omega)=(e^{\beta \omega }+1)^{-1}$ and
$N(\omega)=(e^{\beta \omega }-1)^{-1}$.
In (\ref{H}) we have introduced an effective spectral function
$\chi_{{\bf k},{\bf q}}(\omega)$ as
\begin{equation}
\chi_{{\bf k},{\bf q}}(\omega)=-\frac{1}{\pi}
\sum_{\alpha \beta}g_{\alpha \beta}({\bf k},{\bf q})
\text{Im}D_{\alpha \beta}({\bf q},\omega)
,
\label{chi}
\end{equation}
for spin fluctuations coupled to a particular hole ${\bf k}$--state.

For the single--hole problem at zero temperature \cite{KLR,MH}
the self--energy (\ref{H}) is reduced and only the  contribution due
to $D_{11}({\bf q},\omega)$ remains. Considering the case of
finite hole concentration $\delta$, the full form (\ref{H})
is adopted here, in contrast to Ref.\ \onlinecite{Kit}.
In a proper analysis, Eqs.\ (\ref{G})--(\ref{H}) should be treated
self--consistently, the problem which, to our knowledge, can be
solved only numerically.
Here, we are interested in a particular question of calculating
the damping of QP hole states
in an analytical way based on the well--established
numerical results of SCBA.

Actually, spectral characteristics of a hole propagating in an AFM
background at low level of doping have been investigated
in many works \cite{Rev}. Those results led to the consensus that the
hole spectrum involves a narrow QP band of coherent states
at low energies and a broad continuum of incoherent states.
The corresponding spectral function is then represented as
$\rho_{{\bf k}}( \omega)=\rho^{\text{coh}}_{{\bf k}}(\omega)+
\rho^{\text{inc}}_{{\bf k}}(\omega)$ with
\begin{equation}
\rho^{\text{coh}}_{{\bf k}}(\omega)=Z_{{\bf k}}\delta
(\omega+\mu-E_{{\bf k}})
.
\label{coh}
\end{equation}
The QP weight $Z_{{\bf k}}$ and
the bandwidth $W$ are estimated to be
$Z_{{\bf k}}\simeq J/t\equiv Z$ and $W\simeq 2J$.
The QP dispersion
$E_{{\bf k}}$ in the vicinity of its minima
${\bf k}_{i}=(\pm \pi /2, \pm \pi /2)$ can be expanded as
$E_{{\bf k}_{i}+{\bf k}^{\prime}}\simeq
E_{{\bf k}_{i}}+k^{\prime 2}_{\parallel}/2m_{\parallel}
+k^{\prime 2}_{\perp}/2m_{\perp}$. \cite{MH,IF}
Here, $k_{\parallel}^{\prime}$ and $k_{\perp}^{\prime}$
are the component of $ {\bf k}-{\bf k}_{i}$
in the $(1,-1)$ and $(1,1)$ directions in the Brillouin zone (BZ)
for $\mbox{\bf k}_{i}=(\pi/2, \pi/2)$.
For instance, the anisotropy factor $a=m_{\parallel}/m_{\perp}$
is calculated to be $a=6$ for $J=0.3t$.\cite{MH}
This anisotropy can be absorbed by the following transformation
${\bf k}^{\prime}\rightarrow
(ak_{\parallel}^{\prime},k_{\perp}^{\prime}$), which does not change
our final results. Hence, we further consider the case
$m_{\parallel}=m_{\perp}=m(\sim J^{-1})$. \cite{KLR,MH,IF}

From the above results one may expect that the filling of QP
states leads to a four--pocket FS.
Some arguments have also been given in Refs.\ \onlinecite{IF}
and\ \onlinecite{PL} that the fraction of BZ
covered by these pockets at $T=0$
is equal to the hole concentration $\delta$.
This leads to the following estimations for the
Fermi momentum $k_{F}=\sqrt{\pi\delta}$ and the chemical potential
$\mu=E_{{\bf k}_{i}}+k_{F}^{2}/2m$.

The nearly structureless incoherent part
$\rho^{\text{inc}}_{{\bf k}}(\omega)$
is distributed predominantly above the QP band and
can be approximated as \cite{KH}
\begin{equation}
\rho^{\text{inc}}_{{\bf k}}(\omega^{\prime})=(1/2\Gamma)
\theta(|\omega^{\prime}|-J)\theta (2\Gamma-\omega^{\prime})
,
\label{roinc}
\end{equation}
where $2\Gamma\simeq 2zt$ and $\omega^{\prime}$ is measured
from the middle of the QP band.
In
Eq.(\ref{roinc})
the negative energy cutoff
$\omega_{c}\simeq-J-2\Gamma(1-Z_{k})\delta$
is implied. It is provided by the sum rule
$N^{-1}\sum_{{\bf k}}\int\limits
d\omega n(\omega)
\rho_{{\bf k}}(\omega)=\delta$
taken at $T=0$ with the value of the chemical potential
$\mu$ defined above. Numerical analysis \cite{IF,PL}
of the model allows us to conclude that $\omega_{c}$
does not depend on $T$ (for $T\ll J$).

In the above formulated scheme (\ref{G})--(\ref{H}) the
QP damping is due to scattering by spin--waves.
For QP states near FS
 renormalization of low--lying long--wavelength
spin excitations
($\omega\ll \varepsilon_{F}=k_{F}^{2}/2m,\ q\alt 2k_{F}\ll 1$)
is of crucial importance.
This renormalization is due to the coupling of spin--waves to
"particle--hole" pair excitations and
is described by the polarization operator (\ref{P})
which contains three contributions. The first part
$\Pi^{\text{c-c}}({\bf q},\omega)$ is due to the
transitions within the narrow QP band,
when both $\rho_{{\bf k}}(\omega_{1})$
and $\rho_{{\bf k}-{\bf q}}(\omega_{2})$ in Eq.(\ref{P})
are replaced by $\rho^{\text{coh}}$.
The remaining two terms $\Pi^{\text{c-i}}({\bf q},\omega)$
and $\Pi^{\text{i-i}}({\bf q},\omega)$ are
provided by the coherent--incoherent and incoherent--incoherent
transitions.

First, considering $\Pi^{\text{c-c}}({\bf q},\omega)$ we come to
the following expression for small $|{\bf q}|\ll 1$
\begin{eqnarray}
\Pi_{\alpha \beta}^{\text{c-c}}({\bf q},\omega)
&=&\frac{(zt)^{2}}{N}\sum_{i,{\bf k}^{\prime}}^{}Z_{{\bf k}_{i}}^{2}
M_{\alpha}({\bf k}_{i}+{\bf k}^{\prime},{\bf q})
M_{\beta}({\bf k}_{i}+{\bf k}^{\prime},{\bf q})\nonumber \\
&&\times\frac{\left[n(\varepsilon_{{\bf k}^{\prime}- {\bf q}})-
n(\varepsilon_{{\bf k}^{\prime}})\right]}
{\omega-\varepsilon_{{\bf k}^{\prime}} +
\varepsilon_{{\bf k}^{\prime}-{\bf q}}+i\eta}
,
\label{Pcc}
\end{eqnarray}
where
$\varepsilon_{{\bf k}^{\prime}}=(k^{\prime 2}/2m-\varepsilon_{F})$
is the hole energy referred to the Fermi level
$\varepsilon_{F}=k_{F}^{2}/2m$
and the summation over $i$ is due to the presence
of four equivalent minima.
Since for small momentum $|{\bf q}|\ll 1$
the vertex function $M_{1,2}({\bf k},{\bf q})$
is proportional to $\sqrt{q}$, we make an approximation
$M_{1,2}({\bf k}_{i}+{\bf k}^{\prime},{\bf q})\simeq\pm
\tilde M({\bf k}_{i},{\bf q})$, where
\begin{equation}
\tilde M({\bf k},{\bf q})=
2^{-5/4}q^{-1/2}(q_{x}\sin k_{x}+q_{y}\sin k_{y})
,
\label{ver}
\end{equation}
to keep the leading contributions in Eq.(\ref{Pcc}).
This leads to the following relations between the
elements of the polarization operator:
$\Pi_{11}^{\text{c-c}}({\bf q},\omega)=
\Pi_{22}^{\text{c-c}}({\bf q},\omega)=
-\Pi_{12}^{\text{c-c}}({\bf q},\omega)
=-\Pi_{21}^{\text{c-c}}({\bf q},\omega)\equiv \Pi_{\bf q}(\omega)$.
We also note that the summation over $i$ in Eq.(\ref{Pcc})
introduces an effective interaction
$\sum_{i}|\tilde M({\bf k}_{i},{\bf q})|^{2}=q/\sqrt{2}$.
Then, for $T=0$ we obtain the following expressions
for the real and imaginary parts of $\Pi_{\bf q}(\omega)$,
\begin{eqnarray}
\text{Re}\Pi_{q}(\omega)=C\Bigl\{
 -q
+\text{sgn}[\eta(q,\omega)]
[\nu(q,\omega)] ^{1/2} \nonumber \\
+\text{sgn}[\eta(q,-\omega)]
[\nu(q,-\omega)] ^{1/2}\Bigr\}
,
\label{RPcoh}
\end{eqnarray}
\begin{eqnarray}
\text{Im}\Pi_{q}(\omega)=C\Bigl\{
[-\nu(q,\omega)] ^{1/2}-[-\nu(q,-\omega)] ^{1/2}\Bigr\}
,
\label{IPcoh}
\end{eqnarray}
where $\eta(q,\omega)=m\omega /q+q/2$,
$\nu(q,\omega)=\eta^{2}(q,\omega)-(k_{F})^{2}$
and $C=4\sqrt{2}mt^{2}Z^{2}/ \pi\sim 4\sqrt{2}J/ \pi$.
The step--$\Theta$--functions insuring the positivity of the
arguments of the square roots are implied
in Eqs.(\ref{RPcoh}) and (\ref{IPcoh}).

For further purposes we fix also the asymptotic,
$\omega \rightarrow 0$, behavior
of $\Pi_{q}(\omega)$ for $q<2k_{F}$:
\begin{eqnarray}
\text{Im}\Pi_{q}(\omega)=0, \;
\text{Re}\Pi_{q}(\omega)=
\frac{C\pi \delta q^{3}}{2m^{2}\omega^{2}}
,
\label{spin}
\end{eqnarray}
for $\omega/q>k_{F}/m$, while for $\omega/q<k_{F}/m$ one has
\begin{eqnarray}
\text{Im}\Pi_{q}(\omega)=
\frac{-2Cm\omega}{\sqrt{(2k_{F})^{2}-q^{2}}}, \;\;
\text{Re}\Pi_{q}(\omega)=-Cq
.
\label{hole}
\end{eqnarray}

The limiting case (\ref{spin}) is important in calculating
of the renormalized spin--wave velocity, which will be
shortly discussed below. The limit (\ref{hole}) corresponds
to the region of the spin fluctuation spectrum, generated
by "particle--hole" excitations, which produces  finite
damping of quasiholes near FS. In this respect, we note
the conventional linear $\omega$--dependence of
$\text{Im}\Pi_{q}(\omega)$ in Eq.\ (\ref{hole}) in contrast
to a marginal ($q$-- and $\omega$--independent, at $T=0$)
form of $\text{Im}\Pi_{q}(\omega)$ in Ref.\ \onlinecite{Kit}
(see Eq.(19) there). The reason of this difference can be
explained as follows. Due to the incorrectly defined limits of
integration in Eq.(14) of Ref.\ \onlinecite{Kit} the
authors lost part of the polarization operator (which
is presented by the second term in the curly brackets
in Eq.\ (\ref{IPcoh}) in this paper), which led
to wrong subsequent approximations.

Considering the remaining contributions
$\Pi_{\alpha \beta}^{\text{c-i}}({\bf q},\omega)$ and
$\Pi_{\alpha \beta}^{\text{i-i}}({\bf q},\omega)$ we note that
each of them is characterized by
a threshold energy $\Delta$ for creating a
"particle--hole" pair excitation. Namely,
$\Delta=\varepsilon_{F}$ for processes involved in
$\Pi_{\alpha \beta}^{\text{c-i}}({\bf q},\omega)$
and $\Delta=2J$ for $\Pi_{\alpha \beta}^{\text{i-i}}({\bf q},\omega)$.
Therefore, $\text{Im}\Pi_{\alpha\beta}^{\text{inc}}=0$, where
$\Pi_{\alpha\beta}^{\text{inc}}=\Pi_{\alpha\beta}^{\text{c-i}}+
\Pi_{\alpha\beta}^{\text{i-i}}$, for frequencies
 $\omega < \varepsilon_{F}$ we are interested in.
Evaluation of $\text{Re}\Pi_{\alpha\beta}^{\text{inc}}$
requires the summation over all
virtual processes,which gives a finite estimate for these
quantities even as $\omega\rightarrow 0$.
To the lowest order in $q$ and $\delta$ for
$\text{Re}\Pi_{\alpha\beta}^{\text{inc}}$ we obtain
 $\text{Re}\Pi_{11,22}^{\text{inc}}(q,\omega)\simeq
 -\text{Re}\Pi_{12,21}^{\text{inc}}(q,\omega)\simeq
 -Aq\delta$
where $A$ is positive and can be estimated as
$A\simeq t/\sqrt{2}\{\ln (zt/J)+z^{2}(1-Z_{k})
[\ln (2J/zt\delta)+1]\}$.

A position of the pole in the spin--wave GF in the
long--wavelength limit is now determined as
 \begin{equation}
\tilde{\omega}_{{\bf q}} =\omega_{{\bf q}}\sqrt{1-
 2[Aq\delta-
 \text{Re}\Pi_{q}(\omega)]/\omega_{q}}+O(\delta^{2})
 .
 \label{u}
 \end{equation}
Since the unrenormalized spin--wave velocity $u=\sqrt{2}J$
is much larger than the Fermi velocity $v_{F}=\sqrt{\pi\delta}J$
one has to take in Eq.\ (\ref{u}) the limit (\ref{spin})
which gives $\text{Re}\Pi_{q}(\omega_{q})\simeq\omega_{q}\delta$.
Thus the renormalized spin--wave velocity now reads
$\tilde{u}(\delta)=u\sqrt{1-2(A/u-1)\delta}$.
For the actual values of $\delta$ it holds $A/u\gg 1$ and hence,
one obtains, in accordance with Refs.\ \onlinecite{IF}
and\ \onlinecite{KH} a spin--wave softening due to the presence
of the incoherent part in the hole spectrum.

The above estimation for $\tilde{u}$ is valid up to the critical hole
concentration $\delta_{c}$ which is defined as
$\tilde{u}(\delta_{c})=v_{F}(=\sqrt{\pi\delta_{c}}J)$. In particular, for
$J/t=0.3$ we estimated $\delta_{c}\simeq0.04$. For higher concentrations
$\delta>\delta_{c}$, by taking the corresponding limit (\ref{hole})
one can see from Eq.(\ref{u}) that the pole
$\tilde{\omega}_{{\bf q}}$ becomes purely imaginary.
So, the long--wavelength magnons, with $q\alt 2k_{F}$
lose their identity and can not be now detached from the
incoherent part of the spectrum produced by pair excitations.
Disappearance of the long--wavelength magnons due to dilution
of the AFM state with holes was connected by several authors
\cite{SH} with the occurrence of a phase transition into
a disordered magnetic phase. The applicability of the spin--polaron
model in the disordered phase will be discussed bellow.

Let us consider the effective spectral function (\ref{chi}). By
using the approximated vertex function (\ref{ver}) for the
momentum near FS, $k^{\prime}\sim k_{F}$ $({\bf k}^{\prime}
={\bf k}-{\bf k}_{i})$, we obtain
\begin{equation}
\chi_{{\bf k}_{i}+{\bf k}^{\prime},{\bf q}}(\omega)\simeq
-1/\pi|\tilde{M}({\bf k}_{i},{\bf q})|^{2}
\text{Im}\tilde{D}({\bf q},\omega)
,
\label{chi1}
\end{equation}
where $\tilde{M}({\bf k}_{i},{\bf q})=2^{-3/4}q^{1/2}
\hat{{\bf q}}\hat{{\bf k}}_{i}$ and
$\tilde{D}({\bf q},\omega)=D_{11}-2D_{12}-D_{22}$.
Taking into account the relation between the elements of the polarization
operator $\Pi({\bf q},\omega)$, that has been obtained above, we write
\begin{equation}
\mbox{Im}\tilde{D}({\bf q},\omega)=
4\omega_{{\bf q}}^{2}
\text{Im}\Pi_{q}(\omega)
/|{\cal D}_{{\bf q}}(\omega)|^{2}
.
\label{D}
\end{equation}

For the actual region of the $\omega$-- and $q$--variables,
defined as $\omega/q<v_{F}$, where the part of the spin fluctuation
spectrum responsible for the quasihole damping is located, one has
$|{\cal D}_{{\bf q}}(\omega)|^{2}=[\omega^{2}+c^{2}q^{2}]^{2}
+[2\omega_{{\bf q}}\text{Im}\Pi_{q}(\omega)]^{2}$
with $c=u\sqrt{2(A+C)/u-1}\gg v_{F}$.
This strong inequality allows us to take the static limit,
$\omega\rightarrow 0$,
for ${\cal D}_{{\bf q}}(\omega)$ in Eq.\ (\ref{D}) that results in
\begin{equation}
\chi_{{\bf k}_{i}+{\bf k}^{\prime},{\bf q}}(\omega)\simeq
-1/\pi\sqrt{2}(u/c^{2})^{2}\text{Im}\Pi_{q}(\omega)
(\hat{{\bf q}}\hat{{\bf k}}_{i})^{2}q^{-1}
.
\label{chi2}
\end{equation}
Inserting (\ref{chi2}) into (\ref{H}) one obtains for the
imaginary part of the hole self--energy
\begin{eqnarray*}
\text{Im}\Sigma({\bf k}^{\prime},\varepsilon)\propto
\!\! \int\!\!\!\!\int\!
dq\cos^{2}\!\theta
d \theta
\!\!\int\limits_{0}^{\varepsilon}\!\!
d\omega
\text{Im}\Pi_{q}(\omega)\delta
(\varepsilon-\omega-\varepsilon_{{\bf k}^{\prime}-{\bf q}})\nonumber
,
\end{eqnarray*}
where $ \cos\theta=\hat{{\bf q}}\hat{{\bf k}}_{i}$,
${\bf k}^{\prime}={\bf k}-{\bf k}_{i}$,
$\varepsilon$ is the hole energy referred to the Fermi level
and $\text{Im}\Pi_{q}(\omega)$ is defined in Eq.\ (\ref{hole}).

Like in the conventional considerations of the
2--dimensional (2D) FL, \cite{2D}
the major contribution to $\text{Im}\Sigma$
is given by scattering processes with the momentum transfer ${\bf q}$
almost parallel to ${\bf k}^{\prime}$.
These processes result in the following dependence for $\text{Im}\Sigma$
familiar for the 2D FL:\cite{2D}
\begin{equation}
\text{Im}\Sigma({\bf k}^{\prime},\varepsilon)\propto
f(|\hat{{\bf k}}^{\prime}\hat{{\bf k}}_{i}|)
(\varepsilon^{2}/ \varepsilon_{F})\ln(\varepsilon/ \varepsilon_{F})
.
\label{dam}
\end{equation}
Here, a ${\bf k}^{\prime}$--dependence of $\text{Im}\Sigma$
is due to the anisotropy of the vertex (\ref{ver}) and is given
by $f(|\hat{{\bf k}}^{\prime}\hat{{\bf k}}_{i}|)$
which is a positively defined
smooth function of its variable.

Up to now we have considered the scattering processes
retaining a quasihole in the vicinity of the same hole--pocket.
There exist, however, processes in which the hole scatters
from a given hole--pocket to the opposite or neighboring one,
with momentum transfer
${\bf q}\sim {\bf Q}$ (${\bf Q}$ is the AFM wave vector)
and ${\bf q}\sim {\bf Q}^{\prime}=(\pi,0)$,
respectively. Since the symmetry of the problem provides the
equivalence of $q\ll 1$ and $q^{\prime}=|{\bf Q}-{\bf q}|\ll 1$,
the first process gives just an extra factor 2 in $\mbox{Im}\Sigma$.
Further, the vertex function $\tilde{M}({\bf k}_{i},{\bf q})$
falls much faster at ${\bf q}\simeq {\bf Q}^{\prime}$
than at ${\bf q}\simeq 0$ (or ${\bf Q}$) and
the second kind of processes gives higher order
corrections in $\omega$.
So, the conventional 2D FL behavior is expected for quasiholes
at low doping.

Being originally formulated for a state with an AFM ordered
spin subsystem, the spin--polaron model requires some justification
if one tries to extend it to a disordered phase, {\it i.e.}
either to $\delta > \delta_{c}$
at $T=0$ or $T>0$.
Actually, \cite{Dag} the hole spectrum is weakly affected
by the absence of the long--range order, provided that
the AFM correlations with the radius $\xi \gg R_{p}$ survive
($R_{p}$ is the size of the
spin--polaron associated with a hole).
It means that hole propagation over the same sublattice dominates
and the four--pocket FS survives as well.

Connecting the magnetic phase transition at $\delta=\delta_{c}$
with the disappearance of long--wavelength magnons with
$q\alt 2k_{F}$, we did not find, however, any abrupt change in that
low--lying part of spin fluctuation spectrum which is
responsible for the quasihole
damping. Therefore, one may expect that not only the QP
dispersion relation but also the character of the
quasihole damping (\ref{dam}) do
not change for $\delta$ slightly above $\delta_{c}$.

This picture breaks down with further dilution of the
magnetic subsystem, when the magnetic correlation length becomes
comparable with the size of the spin--polaron. In this case,
the nearest--neighbor hole hopping becomes dominant and
a transition to a large FS takes place.
However, this regime is beyond the scope of the present
consideration.

Considering a possible effect of finite $T$ (low enough to provide
$\xi \gg R_{p}$ ) we point out the existence of a characteristic
temperature $T_{d}(\delta)$ above which one may expect different
behavior of a quasihole subsystem as compared to the low
temperature case, $T \ll T_{d}(\delta)$.
Actually, the Fermi--ensemble of quasiholes goes over into the
strongly nondegenarate regime when the temperature
$T_{d}(\delta)\approx \varepsilon_{F}\approx 1.5J\delta$ is reached
(for instance, $T_{d}\approx 100$K at $\delta=0.05$ for $J=1500$K).
 That is a result of the strong renormalization of the
chemical potential $\mu$ with $T$, which is naturally inherent in
a fermion system of low density. Really, our analytical estimations,
as well as numerical calculations,\cite{PL} show that $\mu$ crosses
the bottom of the QP band at $T\simeq T_{d}$ and for $T>T_{d}$
lies in the low energy incoherent part of the hole spectrum.
This results in a dramatic change in the momentum distribution
function $N(k)$: the four--pocket structure existing at $T<T_{d}$
is almost washed out at $T>T_{d}$.\cite{PL}
 The onset of this strongly nondegenerate regime for hole carriers
should manifest itself in a strong change of the thermodynamic and
transport properties of the system, the problem which requires
further theoretical and experimental studies.

In summary, we have investigated the quasihole damping in the low
doping regime, $\delta \ll 1$, of the 2D $t-J$ model.
The self--energy parts for the hole and magnon GF are derived
within the self--consistent Born approximation.
Based on the well established results \cite{Rev} for the spectral
density function of a hole moving in the AFM background, we first have
calculated renormalization of spin--wave excitations due to the
presence of holes. With increasing hole concentration $\delta$,
softening of the long--wavelength spin--waves followed by their
overdamping at $\delta> \delta_{c}(\approx 0.04)$ has been obtained.
The renormalized spectrum of spin excitations was incorporated
to calculate the imaginary part $\text{Im}\Sigma ({\bf k},\varepsilon)$
of the hole self--energy. It has been shown that $\text{Im}\Sigma$,
as $\varepsilon\rightarrow 0$, possesses the form (\ref{dam})
characteristic of the conventional 2D FL.\\ \\

We would like to thank Prof. N. Plakida, Dr. P. Horsch and
Dr. V. Kabanov for useful discussions and comments.
Financial support by the Russian
Foundation for Fundamental Researches Grant No 96--02--17527, and
the INTAS--RFBR Program Grant No 95--591 is acknowledged.
One of the authors (V. Yu.) acknowledges also the support by NREL
in the framework of Subcontract No AAX--6--16763--01.

%%%%%%%%%%%%%%%%%%%%%%%%%%%%%%%%%%%%%%%%%%%%%%%%%%%%%%%%%%%%%%%%%%%

\end{document}